\documentclass[preprint]{revtex4}

\usepackage{amsfonts,xfrac}
\usepackage{amsmath}
\usepackage{amssymb}
\usepackage{graphicx}
\usepackage{dcolumn}
\usepackage{dsfont}
\usepackage{times}
\usepackage{epsfig,color}
\usepackage[]{units}
\usepackage{soul}
\usepackage{bm}
\usepackage[hang,nooneline]{subfigure}                         
\newcounter{fig}   

\def\epsilon{\varepsilon}

\begin{document}

\title{\textsf{Data-driven prediction and analysis of \\ chaotic origami dynamics}}
\author{Hiromi Yasuda$^{1,2}$, Koshiro Yamaguchi$^1$, Yasuhiro Miyazawa$^1$, Richard Wiebe$^3$, Jordan R. Raney$^2$, and Jinkyu Yang$^1$}
\affiliation{$^1$Department of Aeronautics \& Astronautics, University of Washington, Seattle, WA 98195-2400, USA\\ $^2$Department of Mechanical Engineering and Applied Mechanics, University of Pennsylvania, Philadelphia, PA 19104, USA\\ $^3$Department of Civil and Environmental Engineering, University of Washington, Seattle, WA 98195-2400}

\begin{abstract}
Advances in machine learning have revolutionized capabilities in applications ranging from natural language processing to marketing to health care. Here, we demonstrate the efficacy of machine learning in predicting chaotic behavior in complex nonlinear mechanical systems. Specifically, we use quasi-recurrent neural networks to predict extremely chaotic time series data obtained from multistable origami systems. Additionally, while machine learning is often viewed as a ``black box'', in this study we conduct hidden layer analysis to understand how the neural network can process not only periodic, but also chaotic data in an accurate manner. 
Also, our approach shows its effectiveness in characterizing and predicting chaotic dynamics in a noisy environment of vibrations without relying on a mathematical model of origami systems. Therefore, our method is fully data-driven and has the potential to be used for complex scenarios, such as the nonlinear dynamics of thin-walled structures and biological membrane systems.
\end{abstract}

\maketitle




\clearpage


Chaos has been widely studied for decades in physics, mathematics, and engineering~\cite{strogatz_nonlinear_2015, wiggins_introduction_2003, wiebe_heuristic_2012, dieci_numerical_2011}.
Since chaos is generally defined for deterministic systems as a high sensitivity to initial conditions, it could be considered predictable in a mathematical sense, assuming all relevant information about the system is known.
 In practical terms, however, it is extremely difficult to accomplish this task hindered by unknown factors such as noise and interactions with the surrounding environment. Therefore, it remains a formidable challenge to predict chaotic behavior in practice.

 Here, we study a data-driven approach to analyze chaotic time series data and predict future response by using machine learning techniques. 
Recurrent neural networks (RNNs) constitute a powerful machine learning approach for processing and predicting time-series data~\cite{Connor1994,Husken2003} (see Fig.~S1a for schematic illustration of a standard RNN).
Due to such capabilities, RNNs have been recently employed for physics problems~\cite{Hughes2019,Pathak2017,Pathak2018,Vlachas2018,Yeo2019}.
  However, RNNs are usually considered to be a ``black box'' for learning and predicting  time series. 
  Thus, interpretation of the neurons' processing a time series, especially for chaotic data, has remained elusive.
  This is partly because the activation for each neurons in an RNN for the current time step depends on the activation of every other neuron at the previous time step (denoted by a red circle in Fig.~S1a).
  This deep level of coupling interaction makes it challenging to extract meaningful information about the effects of individual neurons.

Recently, quasi-recurrent neural networks (QRNNs) have been developed, particularly for natural language analysis (see Supplementary Note 1 and Fig.~S1b)~\cite{Bradbury2017}. They exhibit faster processing of time series data and competitive performance compared with other RNNs. Most notably, QRNNs allow access to hidden layers, offering the ability to scope the data set and assign ``meaning'' to each neuron. 
 The QRNN is composed of convolutional layers which process time-series data in parallel across each time step, and pooling layers in which recurrent relations can be implemented (Fig.~S1b).
 Due to the element-wise calculation in the pooling function, activation of each neuron does not depend on the past outputs of other neurons. 

 In this study, we demonstrate not only prediction of both periodic and chaotic data, but also analysis of hidden units' distinctive responses to such dynamic conditions by using the QRNN.
 To examine effectiveness of our approach, especially in the experimental context, we need to produce unique sets of dynamic data containing periodic, subharmonic, and chaotic trajectories in a controllable manner. 
 One of the examples showing such behavior is a structure with intrinsic bistability~\cite{Tseng1971,Moon1979,Harne2013}. In this study, we design and fabricate a versatile bistable mechanical system based on origami unit cells, specifically triangulated cylindrical origami (TCO)~\cite{Kresling1995} as shown in Fig.~\ref{fig:Bistable}a.
 These TCO cells, in serial connection, can provide highly tunable properties~\cite{Jianguo2015,Yasuda2017,Yasuda2019} as well as multi-degree-of-freedom nature, thus providing an ideal playground to examine the effectiveness of our data-driven prediction.
 
Based on the experimentally measured time-series data, we explore the feasibility of extracting meaningful system information from our data-driven approach. Prior research also attempted to find  parameters/functions of the governing equations by using data-driven approach~\cite{Rudy2017,Raissi2019,Champion2019,Iten2020}.
However, our study differs from these previous work in that no mathematical model of the system or even the nature of dynamics (e.g., definition of chaos) is provided. In our analysis of the hidden layers of the QRNN, we find that the neuron activation of specific hidden units behaves sensitively to the input data, which naturally enables the QRNN to distinguish between chaotic and non-chaotic origami dynamics.
This approach has great potential for predicting complex dynamics of structures by allowing access to evolutionary steps and providing more parameters to reach reliable decisions.
\section*{Experimental demonstration of chaotic behavior of the Triangulated Cylindrical Origami}
 
\begin{figure*} 
\centering
\includegraphics[width=1.\linewidth]{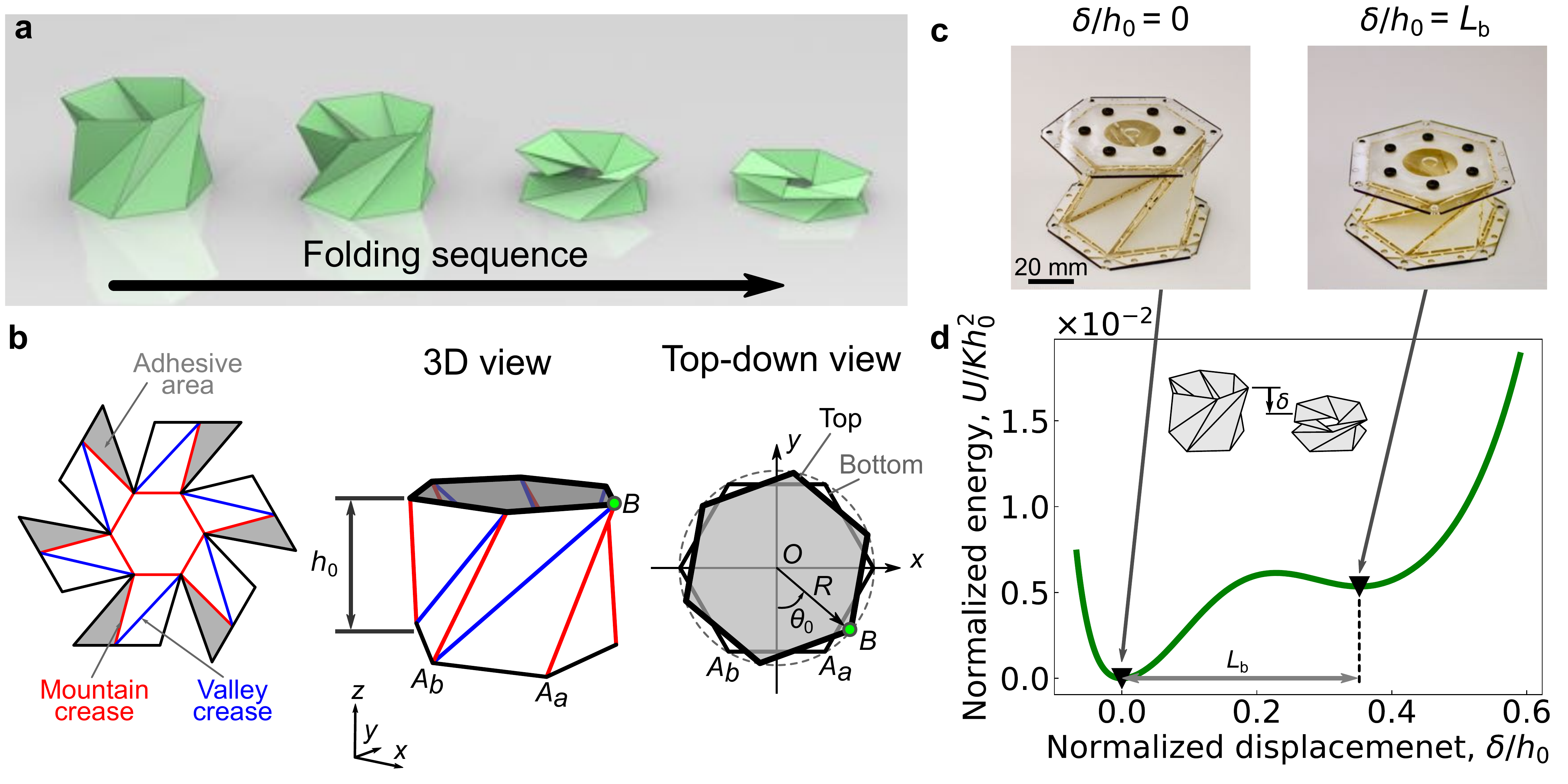}
\caption{\textbf{Folding behavior of the triangulated cylindrical origami (TCO) unit cells.} (\textbf{a}) Folding sequence of the TCO with $(h_0, \theta_0, R) = (50, 70^\circ, 36)$. (\textbf{b}) The flat sheet with crease patterns of the TCO (\textit{Left}) is folded into a 3D cylindrical shape (\textit{Center}). The parameters to define the initial shape are the initial height ($h_0$), initial rotational angle ($\theta_0$), and radius of the cross-section ($R$)  as shown in the top-down view (\textit{Right}).  (\textbf{c}) Actual prototype of the TCO unit cells with $h_0=50$ mm in its initial state (\textit{Left}) and second stable configuration (\textit{Right}). 
(\textbf{c}) Energy landscape calculated from static experiments on the paper prototypes show the bistable behavior, i.e. there exists two energy minima denoted by the triangle markers.} 
\label{fig:Bistable}
\end{figure*}

  Origami has been extensively studied recently due to its tailorable static responses, and we show that a unique TCO-based platform can produce rich data sets from its complex dynamics, especially chaos, thus enabling to examine the effectiveness of our data-driven approach.
  One of the interesting features of the TCO is that its axial and rotational motions are coupled with each other (see Fig.~\ref{fig:Bistable}a where the folding sequence of the TCO unit cell is depicted, and also Supplementary Movie 1 for folding animation).
 Figure~\ref{fig:Bistable}b (\textit{Left}) shows the flat sheet with crease patterns of the TCO and its folded shape.
 To describe the initial shape of the TCO unit cell, we define the initial height ($h_0$), initial rotational angle ($\theta_0$), and radius of the cross-section ($R$)  as shown in Fig.~\ref{fig:Bistable}b (\textit{Center}--\textit{Right}). 
 
 To analyze the potential energy, we fabricate prototypes by using construction paper sheets cut by a laser cutting machine (see the Method; Supplementary Movie 2).
 Figure~\ref{fig:Bistable}c shows our fabricated prototype of the TCO unit cell 
 with $(h_0, \theta_0, R)=(50$ mm, $70^\circ$, 36 mm$)$.
 We conduct quasi-static 
 cyclic loading tests on this paper prototype (see Fig.~S2) to examine and enhance the repeatability of folding/unfolding behavior.
 After this cyclic loading, we extract the force-displacement relationship from compression tests~\cite{Yasuda2017} and obtain the energy curve as shown in Fig.~\ref{fig:Bistable}d.
Here, the energy is normalized by the initial height ($h_0$) and stiffness ($K$) 
at the initial unstretched state, i.e., $\delta=0$ (Fig.~S2c).
 We observe the bistable behavior, such that the TCO unit cell possesses two local minima in its energy landscape where the distance between these two stable states ($L_{b}$) is 0.35.
 This characteristic distance will be used to aid in analyzing the dynamic behavior of the TCO.

\begin{figure}
\centering
\includegraphics[width=0.5\linewidth]{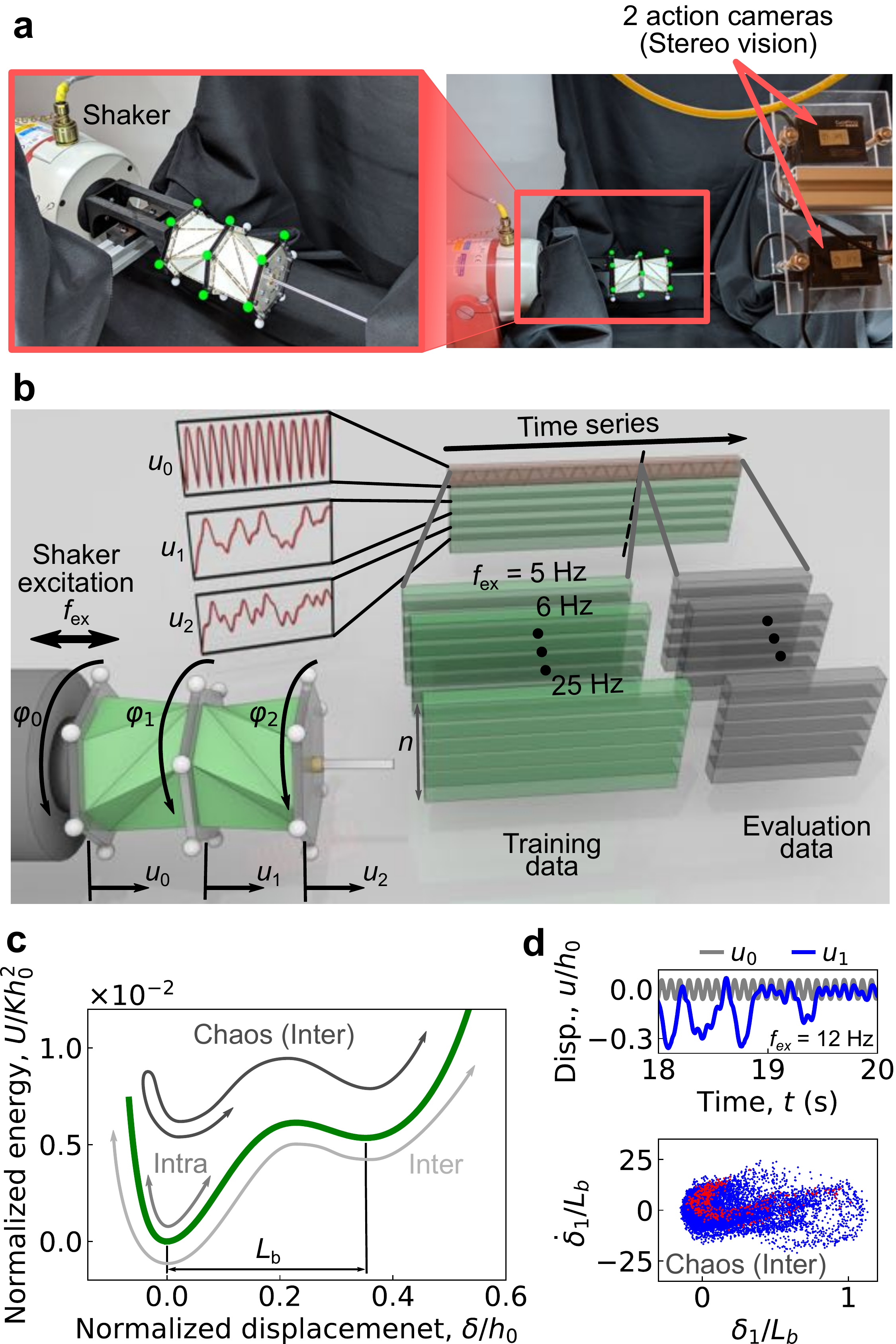}
\caption{ \textbf{Dynamic testing on the two-TCO structure.} (\textbf{a}) Actual testing set up for the vibration test. The structure consists of two TCO unit cells with $(h_0,\theta_0,R)=(50$ mm$, \pm70^\circ, $36 mm$ )$. The input excitation is applied by a shaker, and the folding motion of these two TCO unit cells is captured by two action cameras, together with a customized DIC program. (\textbf{b}) Conceptual illustration of operation on the measurement data for training and evaluating the QRNN. From the DIC, we obtain displacement ($u_i$) and rotational angle ($\varphi_i$) where $i=0,1,2$. The measured data is separated into two data sets; Training data and evaluation data. The experiment was conducted for different excitation frequency ranging from 5 Hz to 25 Hz. (\textbf{c}) Schematic illustration of classifying different folding behavior of the bistable TCO unit cell with double-well potential energy. (\textbf{d}) Measurement results for the excitation frequency of 12 Hz. The displacement of the left-most section ($u_0$), which is attached to the shaker, shows sinusoidal waves, while $u_1$ indicates chaotic motion (\textit{Top}). Let $\delta_i=u_{i-1}-u_i$, we plot a phase portrait in which blue dots are all measurement data and red dots represent Poincar\'{e} map (\textit{Bottom}). }
\label{fig:DynamicTest}
\end{figure}

Forced dynamic tests of a system of two connected TCO unit cells were used to create the chaotic response data sets.
 The unit cells had properties $(h_0,\theta_0,R)=(50$ mm$, \pm70^\circ, $36 mm$ )$, with the left-most cross-section attached to a shaker that generated harmonic excitation (see Fig.~\ref{fig:DynamicTest}a; the Method). 
 The experiment was conducted for different excitation frequency ranging from 5 Hz to 25 Hz.
 The folding motion of these two TCO unit cells is measured by two action cameras together with a customized digital image correlation program.
 In the experiments, we measured displacement ($u_i$) and rotational angle ($\varphi_i$) of each cross-section ($i=0,1,2$, see Fig.~\ref{fig:DynamicTest}b for the notation).
 We use these measured data as well as the velocities, $\dot u_i$ and $\dot \varphi_i$, numerically calculated from $u_i(t)$ and $\varphi_i(t)$.
  This measured data is separated into two data sets, the first of which is used for training, and the second for evaluation (Fig.~\ref{fig:DynamicTest}b).

A schematic diagram of the dynamic folding behavior is shown in Fig.~\ref{fig:DynamicTest}c overlaid on the underlying double-well potential energy landscape.
 We can define three different regimes~\cite{Harne2013}; Intrawell, Interwell (Periodic), and Interwell (Chaotic) vibrations as shown in Fig.~\ref{fig:DynamicTest}c.
 The intrawell oscillation means that the system exhibits small oscillations about one of the two local potential minima. If the system overcomes the energy barrier and goes to the other stable state, we observe the interwell vibrations, which is typically either periodic or chaotic. Quasi-periodic responses may also occur in nonlinear systems, though that was not observed in the origami experiments. 
 
 Figure~\ref{fig:DynamicTest}d (\textit{Top)} shows the measurement results for an excitation frequency ($f_{ex}$) of 12 Hz. The displacement of the left-most section ($u_0$), which is attached to the shaker, shows input sinusoidal waves. However, $u_1$ indicates chaotic motion. 
 Here we define $\delta_i=u_{i-1}-u_i$ and plot a phase plane for $\delta_1$ in which blue dots are all measurement data and red dots represent Poincar\'{e} map~\cite{Virgin2000} as shown in Fig.~\ref{fig:DynamicTest}d (\textit{Bottom)}.
 The horizontal axis is normalized by the distance between two local minimum points ($L_{b}$). Therefore $\delta_i/L_{b}=1$ indicates that the TCO unit cell transits to the second stable regime.
 It should be noted, however, that it is possible for a forced dynamic response to transit over the unstable potential hilltop, and reverse course without ever reaching the stable potential minimum. Thus, this normalization provides only a nominal indicator of a completed transit between the two stable equilibria. 
 The experiment result for $f_{ex}=12$ Hz clearly shows that the vibrations take place not only around the first stable state, but also around the other energy local minimum state aperiodically, which corresponds to the chaotic interwell vibration. This manifests the capability of our TCO system to form chaotic dynamics (see Supplementary Movie 3 for experimental measurements at different excitation frequencies). 

\section*{Prediction based on quasi-recurrent neural networks}
 Based on the experimentally measured data, we predict whether a response is chaotic or periodic by employing the QRNN technique~\cite{Bradbury2017} (see Supplementary Note 1 for the detail information about the QRNN).
 This prediction relies solely on the data obtained from the experiment, and therefore, prior knowledge about a mathematical model of the system is not required.
 To predict the chaotic/periodic folding motion of the TCO structure, we use the QRNN consisting of three hidden layers. Each layer is composed of 352 units.
 The input data $\mathbf{X}$ contains $n=12$ components ($u_i, \dot u_i, \varphi_i, \dot \varphi_i$ for three different cross-sections), and each component has $T=128$ data points, i.e., time steps, which correspond to data length of 0.53 s given the action camera's sampling frequency of 240 fps.
  Using the input data of $\mathbf{X} \in {{\mathbb{R}}^{128 \times 12}}$ from $t_1$ to $t_{128}$, the QRNN predicts all 12 variables for next 32 time steps.
  The total duration of the measured data contains 8000 time steps (33.3 s), and we use first 5600 time steps (23.3 s) for training and the other 2400 time steps (10.0 s) for evaluating the prediction. We obtain 21 sets of such time-series data for 21 frequency steps (i.e., $f_{ex}=5,6,7,...,25$ Hz, see Fig.~\ref{fig:DynamicTest}b for the schematic illustration of our data sets composed of training and evaluation data in various frequencies). 
  We run the training for 100 epochs (see the Method for the parameters used for the QRNN training; also see Fig.~S3 for the training results).
  Please note that we train one QRNN system by using all frequency cases, which enables to predict both periodic and chaotic cases, instead of training neural networks for a specific frequency and predict the dynamics at the corresponding frequency case.

\begin{figure}
\centering
\includegraphics[width=1.\linewidth]{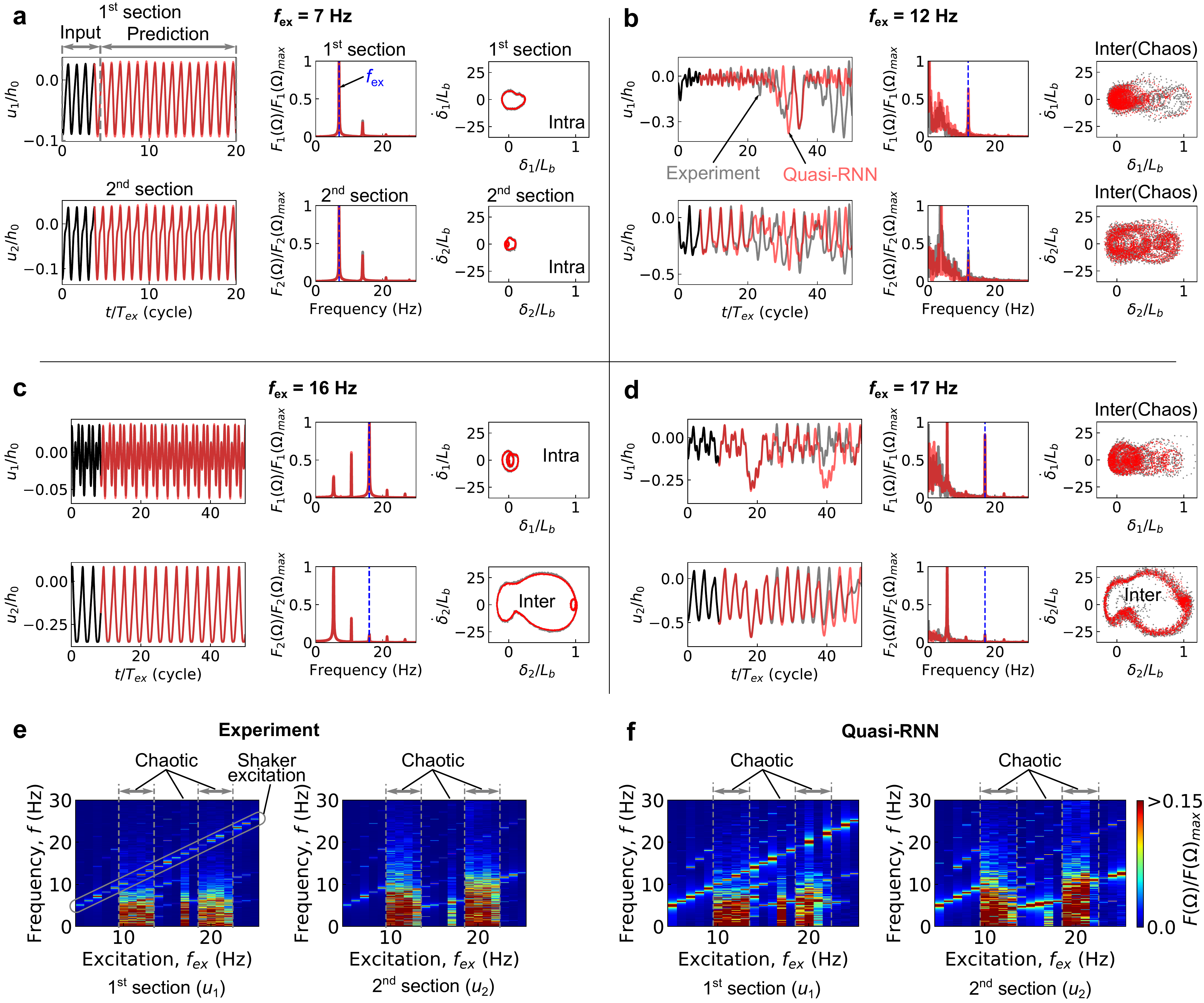}
\caption{ \textbf{QRNN predictions for the folding motion of the TCO system.} (\textbf{a}--\textbf{d}) Predictions from the QRNN (denoted by red color) are compared with the evaluation data from the experiments (denoted by grey color) for different excitation frequencies; $f_{ex}=$ (\textbf{a}) 7 Hz, (\textbf{b}) 12 Hz, (\textbf{c}) 16 Hz, and (\textbf{d}) 17 Hz. In the displacement-time plots (\textit{Left}), displacement is normalized by the initial height ($h_0$) and time is normalized by excitation period ($T_{ex}=1/f_{ex}$). Black solid lines are the initial input data for the QRNN, which are obtained from the evaluation data. To analyze time series data, FFT is applied to $u_1$ and $u_2$ (\textit{Center}). Blue dashed line indicates the excitation frequency ($f_{ex}$). Phase planes show different folding behaviors of each TCO unit cell (\textit{Right}). (\textbf{e}--\textbf{f}) Spectrum analysis for different excitation frequencies from 5 Hz to 25 Hz is applied to $u_1$ (\textit{Left}) and $u_2$ (\textit{Right}). Areas bounded by grey dashed vertical lines indicate chaotic regime, and there is a good agreement between (\textbf{e}) experiments and (\textbf{f}) predictions. }
\label{fig:QRNN_Prediction}
\end{figure}

 Figures~\ref{fig:QRNN_Prediction}a-d show the predictions made by the QRNN (denoted by red color in the figure) compared with the actual data from the measurements (denoted by grey color) for four different excitation frequencies: $f_{ex}=$ 7 Hz (Periodic), 12 Hz (Chaotic), 16 Hz (Periodic), and 17 Hz (Chaotic) (see the Supplementary Movie 4 for the entire folding motions of the TCO unit cells reconstructed from both experiments and predictions).
 In Figs.~\ref{fig:QRNN_Prediction}a-d, left insets show the displacement-time history of $u_1$ and $u_2$, and time is normalized by excitation period ($T_{ex}=1/f_{ex}$).
 The prediction of the QRNN shows excellent agreement with the experimental data for the periodic cases ($f_{ex}=7$ and 16 Hz).
 For the chaotic cases ($f_{ex}=12$ and 17 Hz), the QRNN exhibits quantitatively accurate prediction through approximately 30 excitation cycles, and then the deviation begins growing. It is worth noting that, even with a well-characterized chaotic system (i.e., having access to the governing equations), the sensitivity to initial conditions (and numerical rounding errors) means that quantitative deviation is always expected. Hence, qualitative matching is a more realistic goal. This can be seen by the fact that the frequency spectrum obtained from time series of $u_1$ and $u_2$ shows qualitatively similar trend as shown in the center insets in Fig.~\ref{fig:QRNN_Prediction}a-d. Specifically, the increase of lower frequency components for chaotic responses is successfully captured.
 Additionally, even though the prediction from the QRNN deviated from the experimental results as the number of excitation cycle increases, the QRNN outputs surprisingly reasonable peak deformations of both TCO unit cells. For example, the given TCO configuration with $\theta_0=+70^\circ$ follows the counter-clockwise rotation ($\varphi>0$) under compression and vice versa (i.e., clockwise rotation under tenstion), which is also predicted by the QRNN (see Fig.~S4 for the configuration space of the first and second TCO units as a function of $u$ and $\varphi$). 
 
 The phase portraits, plotted as a function of displacement and velocity, are shown in the right-side insets in Fig.~\ref{fig:QRNN_Prediction}a-d. These show the capability of the QRNN to produce complicated folding behaviors of the TCO structures.
 For instance, lower frequency excitation creates intrawell oscillation for both TCO unit cells, whereas $f_{ex}=16$ Hz triggers 
 intermittent intrawell and interwell oscillations in the same structure. This unique behavior is accurately captured in the prediction.
 In addition to these periodic responses, $f_{ex}=17$ Hz case exhibits interwell chaotic motion of the first unit cell, while the second unit cell shows interwell periodic oscillations. This is also well expressed by the QRNN (see Supplementary Movie 5 for measured and predicted phase portraits for all excitation frequency cases).
 
 To analyze qualitative behaviors for all excitation frequency cases, we performed spectral analysis on different excitation frequencies from 5 Hz to 25 Hz and constructed surface plots as shown in Fig.~\ref{fig:QRNN_Prediction}e-f. 
 In this figure, areas bounded by grey dashed vertical lines indicate chaotic response, and shows good agreement between experiments and predictions (compare surface maps in panel \textbf{e} and \textbf{f}).
 Given the presence of unknown factors in the experiment, the QRNN shows remarkably accurate prediction capability based purely on experimentally-measured data with noise.

\section*{Visualization of the response of the hidden layers to periodic/chaotic data}
\begin{figure*} [t] 
\centering
\includegraphics[width=1.\linewidth]{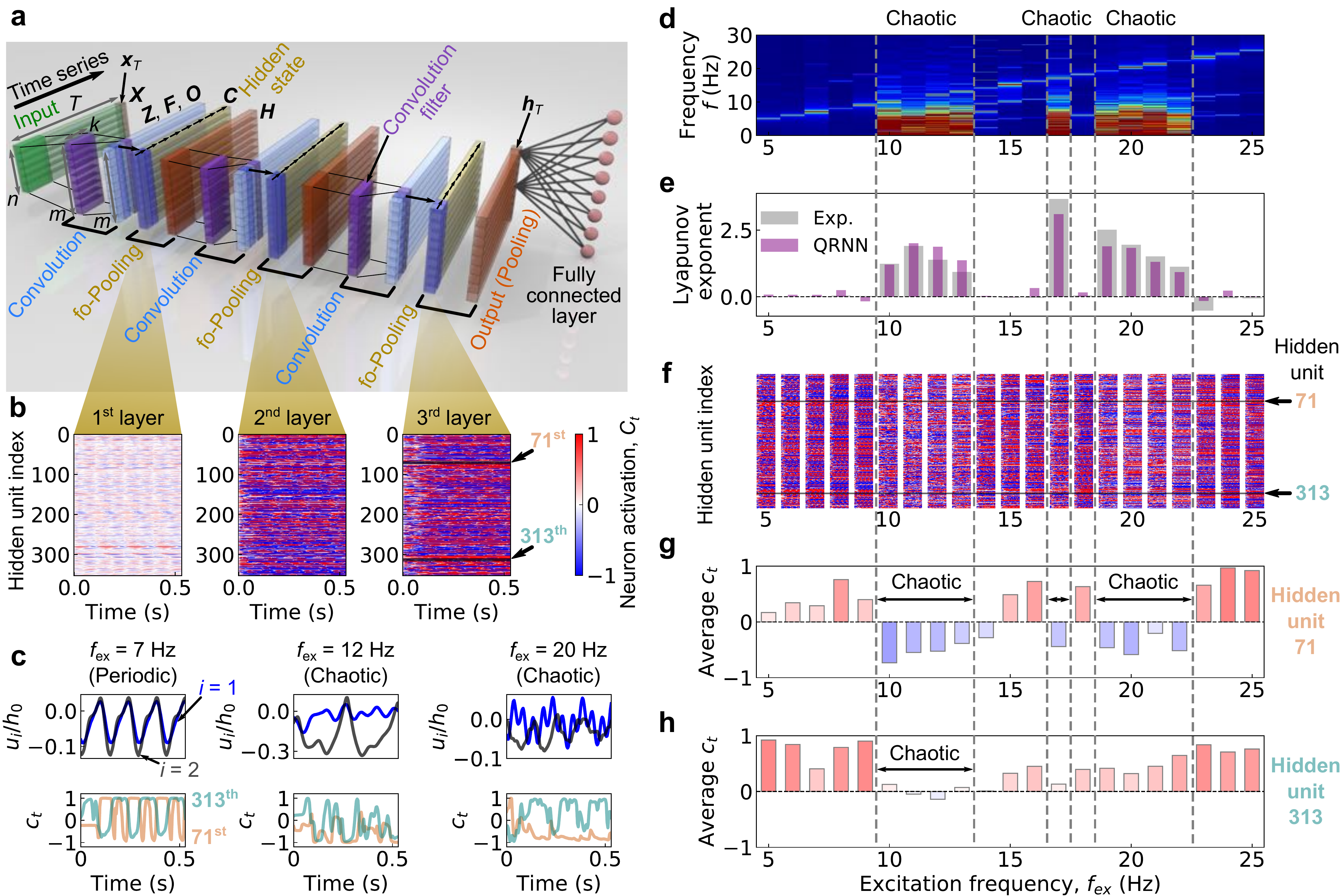}
\caption{ \textbf{Visualization of the hidden states of the QRNN} (\textbf{a}) Schematic illustration of our QRNN configuration composed of three hidden layers is shown. (\textbf{b}) The hidden states of each layer for $f_{ex}$ = 7 Hz (periodic) case are visualized as surface plots where color denotes activation of each hidden unit. (\textbf{c}) To show how the individual hidden units response to the input data, (\textit{Upper}) the displacement-time plots of the initial input data, $u_1$ (blue) and $u_2$ (grey), and (\textit{Lower}) the neuron activation of the 71st (orange) and 313th (green) hidden units are shown. We show the three different excitaion frequency cases: (\textit{Left}) 7, (\textit{Center}) 12, and (\textit{Right}) 20 Hz. To compare all the frequency cases, we perform (\textbf{d}) the FFT analysis on $u_1(t)$ and (\textbf{e}) the Lyapunov exponent calculation, which are compared with the hidden state responses for the third layer as shown in (\textbf{f}). To quantify neuron activation for each excitation frequency, we extract and calculate the average value of neuron activation value $C_t$ for (\textbf{g}) 71st and (\textbf{h}) 313th neurons.}
\label{fig:HiddenStateAnalysis}
\end{figure*}

 This section focuses on the analysis of the hidden-state QRNN computing. Figure~\ref{fig:HiddenStateAnalysis}a shows the schematic illustration of the QRNN structure composed of three hidden layers. 
 Note that the elementwise multiplication in QRNN \textit{fo}-pooling enables the analysis on individual hidden units, because different hidden units do not interact directly in a single pooling layer. This allows independent calculation of each hidden unit until the next QRNN hidden layer. 
For example, the responses of each hidden state to the initial input data at $f_{ex}=7$ Hz are visualized as a function of time and hidden unit index in Fig.~\ref{fig:HiddenStateAnalysis}b.
 The hidden state is composed of $C_t$ vectors with 352 hidden units.
 Thick red/blue colors indicate strong neuron activation, whereas the white color implies reduced activation.
 These visualizations shown several patterns in the hidden states. Extracting the activation history of specific hidden units from the final third hidden layer also shows how a specific neuron processes input data.
 The upper and lower insets of Fig.~\ref{fig:HiddenStateAnalysis}c shows the initial input data ($u_1$ and $u_2$) and the neuron activation of 71st and 313th hidden units, respectively. 
 Interestingly, these hidden units show different neuron activation behaviors between periodic ($f_{ex}=7$ Hz; left inset in Fig.~\ref{fig:HiddenStateAnalysis}c) and chaotic ($f_{ex}=12$ Hz; center inset) cases.
 In addition, in the case of $f_{ex}=20$ Hz (chaotic), we observe notably different behavior between the 71st and 313th hidden units in response to the same input data.
 
 To analyze the neuron activation for different periodic/chaotic regimes, we compare FFT analysis on the experimentally measured data $u_1(t)$ (Fig.~\ref{fig:HiddenStateAnalysis}d) with the Lyapunov exponent calculation (Fig.~\ref{fig:HiddenStateAnalysis}e, see also Supplementary Note 2 for details of the Lyapunov exponent calculation). 
 From the frequency spectrum and the Lyapunov exponent analysis, we identify the three different chaotic regimes: 10-13 Hz, 17 Hz, and 19-22 Hz (bounded by the grey dashed lines in Fig.~\ref{fig:HiddenStateAnalysis}d-h).
 The response of the third hidden layer is  also visualized in Fig.~\ref{fig:HiddenStateAnalysis}f. 
 Based on our observation of the 71st and 313th hidden units above, we extract the neuron activation of those two units and calculate the average activation value over the time length of the input data (i.e., 0.53 s).
  Figures~\ref{fig:HiddenStateAnalysis}g and h show the average activation of 71st and 313th hidden units, respectively.
 The 71st hidden unit shows drastically different behaviors depending on the regimes, i.e., positive (negative) values for periodic (chaotic) cases.
 In addition, it is interesting that 313th hidden unit exhibits weaker activation only for the first two chaotic regimes.
 We confirm this unique response of hidden units from 10 different training runs (see Fig. S5 for different patterns of hidden unit responses, and also the Method for how we find such a hidden unit with unique response).
 
 We observe that the QRNN allows not only the prediction of origami folding behavior, but also the classification of chaotic/periodic input data of the TCO systems by simply monitoring the neuron activation of the hidden units in the final hidden layer.
 The hidden layer analysis can be useful to understand how the QRNN responses to the input data. This simple approach has great potential to characterize and determine chaotic vibrations, providing a reason behind the decision. 
 Note that the concept of chaos is not implemented specifically during training process. Thus, the QRNN itself acquires prediction capability in the process of the training based on the hidden units' distinctive responses to input data between periodic and chaos cases. 

\section*{Discussion}
 In this study, we have demonstrated a data-driven approach to predict and analyze chaotic/periodic behaviors by using QRNN.
 Given challenges of chaos prediction, we have built a mechanical platform composed of origami unit cells and have generated different types of time series data based on intrawell, interwell periodic, and interwell chaotic vibrations.
 By utilizing experimentally measured data, we have trained the QRNN composed of three hidden layers and demonstrated the effectiveness of predicting chaotic/periodic time series.
 One of the unique features of the QRNN based approach is that it allows hidden layer analysis readily without adding extra functions to neural networks. By leveraging this feature, we have calculated average neuron activation of the hidden units in the final layer to examine their response to the input data.
 This simple approach has revealed the different responses of the QRNN's hidden units to the system's dynamic condition, depending on whether it is chaotic or not.
 Also, this approach has potential to provide the reason why the QRNN produces chaotic data, and specifically what parts of the input data lead to different responses of the QRNN.
 Based on these unique features of the QRNN, our approach can contribute to better understanding of both chaos and machine learning techniques for complex dynamic systems. 

\section*{Methods}
\textbf{Prototype fabrication and compression test}
 We used construction paper sheets (Strathmore 500 Series 3-PLY BRISTOL; 0.5 mm paper thickness) for the main origami body and extruded acrylic plate (United States Plastic; 1.6 mm thick) for the interfacial polygon (hexagon in current study), which are both cut by a laser cutting machine (VLS 4.6, Universal Laser Systems).
 Here, folding lines or the crease lines are based on the compliant mechanisms for accurate and consistent folding behavior.
 These are then assembled into TCO unit cells with adhesive sheets (Archival Double Tack Mounting Film, Grafix). See Supplementary Movie 2 for fabrication process.
 Since TCO unit cells are assembled by hand, each unit cell exhibit different force-displacement behavior. This uncertainty in the quality of each prototype significantly influences the repeatability and consistency of the folding/unfolding motion. To avoid this, each unit cell underwent 200 cycles with a controlled displacement from -3 mm (tension) to 15 mm (compression) at 6 mm/s as a preconditioning process~(detailed can be found in Ref.~\cite{Yasuda2019}). 

\textbf{Dynamic test}
 We conducted the dynamic test on a chain composed of two TCO unit cells with $(h_0,\theta_0,R)=(50$ mm$, \pm70^\circ, $36 mm$ )$.
 The left end of the two-TCO-unit system was connected to the shaker (LDS V406 M4-CE, Br\"uel \&  Kj\ae r). 
 The shaker was excited to apply single-frequency harmonic excitation to the system.
 To track the folding/unfolding motion of the TCO unit cells during the dynamic testing, we used two action cameras (GoPro Hero4) and customized Python codes for non-contact digital image correlation.
 We captured images from these two cameras at 240 fps, and identified the coordinates of the spherical markers attached to the corner of the interfacial polygons (see Fig.~\ref{fig:DynamicTest}a) based on the triangulation method~\cite{Yasuda2019}.

\textbf{Prediction based on quasi-recurrent neural networks}
The QRNN used for the prediction consists of three hidden layers, and the filter width ($k$) of 6 is used for each convolution layer (see Fig.~\ref{fig:HiddenStateAnalysis}a).
Training was carried out on minibatches of 50 data sets by using the Adam optimization algorithm~\cite{Kingma2015} with learning rate of 0.001 and decay rate of 0.8.
To create each data set, we randomly chose excitation frequency out of 21 different frequencies ($f_{ex}=5$ Hz $\sim$ 25 Hz), and from the selected frequency data with 5600 time steps, we randomly selected input data with 128 time steps and next 32 times step data to calculate error of prediction.
We implemented this QRNN in custom codes written in Python by using the machine learning library, TensorFlow, and we performed the training/prediction.

\textbf{Analysis on the hidden units}
To compare the activation of the hidden units for different frequencies, we calculated the average value over the duration of the input data (0.53 s).
Let $c_{ave,n}(f_{ex})$ be the average activation of $n$ th hidden unit in the final hidden layer for the excitation frequency of $f_{ex}$, we constructed a vector composed of $C_{ave,n}(f_{ex})$ for different frequencies as $\mathbf{C}_{ave,n}= \left[ c_{ave,n}(5~\text{Hz}), c_{ave,n}(6~\text{Hz}), \cdots, c_{ave,n}(25~\text{Hz}) \right]$, which is visualized as a bar graph as shown in Fig.~\ref{fig:HiddenStateAnalysis}g.
To find unique response of a hidden unit, we examined every hidden units by calculating the difference between $\mathbf{C}_{ave,n}$ and a desired response pattern.
For example, if we look for a hidden unit showing negative activation for chaotic data, we set a vector $\mathbf{C}_{target} = \left[ c_5, c_6, \cdots, c_i, \cdots, c_{25}  \right]$ where
\begin{equation}
{{c}_{i}}=\left\{
\begin{matrix}
  0.5\ \ \ \ \ \ \ \text{(Periodic)} \\ 
 -0.5\ \ \ \ \ \text{(Chaotic)} \\ 
\end{matrix} \right.  \ \ \ \ (i=5,6,\cdots,25).
\end{equation}
Then, we selected the hidden unit which has the minimum value of the norm $\|\mathbf{C}_{ave,n} - \mathbf{C}_{target}\|$.
By changing $\mathbf{C}_{target}$, we can explore the different types of hidden unit activation patterns (see Fig. S5).

\section*{Data Availability}
Data supporting the findings of this study are available from the corresponding author on request.

\begin{acknowledgments}

HY, KY, YM, and JY are grateful for the support of the National Science Foundation (Grant No. CAREER-1553202 and CMMI-1933729) and Washington Research Foundation. HY and JRR gratefully acknowledge support from the Army Research Office award number W911NF-17–1–0147 and Air Force Office of Scientific Research award number FA9550-19-1-0285. KY is supported by the Funai Foundation for Information Technology.
We also thank Professors Nathan Kutz (University of Washington) and Christopher Chong (Bowdoin College) for helpful discussions. 

\end{acknowledgments}

\section*{Author contributions}
H.Y. and R.W. proposed the research; H.Y. and Y.M. conducted the experiments; H.Y. and K.Y. performed the numerical analysis; R.W. and J.Y. provided guidance throughout the research. H.Y., K.Y., R.W., J.R.R, and J.Y. prepared the manuscript.

\section*{Additional information}
The authors declare that they have no competing financial interests. Correspondence and requests for materials should be addressed to Jinkyu Yang~(email: jkyang@aa.washington.edu).

\bibliography{origami.bib}


\newpage

\end{document}